# Bhabha scattering at ILC250

F. Richard

Laboratoire de l'Accélérateur Linéaire (LAL), Centre Scientifique d'Orsay, Université Paris-Sud XI, BP 34, Bâtiment 200, F-91898 Orsay CEDEX, France

______________________________________________________________________

**Abstract**. *This note intends to give an estimate on the sensitivity of the channel ee->ee at the future ILC250. At variance with other two fermion processes, the so-called Bhabha process is influenced by t-channel $Z/\gamma$ exchange. In spite of the complexity of the resulting angular distribution of this process, one observes a good sensitivity to Z' exchange, similar to those observed in annihilation channels. This feature is illustrated within the gauge-Higgs unification model, GHU, which shows an impressive indirect sensitivity to the mass of Z' particles, up to ~20 TeV for the leptonic channels. Beam longitudinal polarisation and high luminosity are the key ingredients for this result. Measuring the Z'ee coupling with the Bhabha process allows to measure separately Z'$\mu\mu$ and Z'$\tau\tau$ coupling, which serves for a precise test of lepton universality. Z'bb and Z'tt couplings show good sensitivities to GHU. LHC and HE-LHC sensitivities are also discussed.*

## I.   Introduction

The future international linear collider (ILC) plans to operate at 250 GeV [1]. This represents a modest increase in energy with respect to LEP2 which has operated up to ~210 GeV, but is a major increase in luminosity. Indeed, the four LEP experiments have only collected 2.5 fb-1 at LEP2, while ILC plans to collect 2000 fb-1, a factor almost 1000 increase. Moreover, ILC will provide an electron beam longitudinally polarised, with an average polarisation of 80%, a crucial ingredient for electroweak analyses. Whether the positron beam will also be polarized still remains uncertain, but this aspect does not play a decisive part in the present analysis.

Due to t-channel photon exchange, Bhabha scattering can be studied with an excellent statistical precision. While it is usually assumed that new physics will mostly affect heavy flavours, top and b quarks and the τ lepton, one should carefully check the electron coupling properties since any anomalous electron coupling will propagate into all other fermionic processes involved at ILC. Obviously the Bhabha channel offers the best opportunity to do so.

In section II, I will go through physics motivations for the present analysis. In section III I recall what was achieved at LEP1 and LEP2. I will then discuss the method to be used which are specific to Bhabha scattering.

In section IV, I will evaluate the sensitivity of the fermionic channels ee, μμ, ττ, bb and tt in a specific model, the gauge-Higgs unification (GHU) model, and compare ILC250 to LEP2 and LHC.

I will show in section VI how it will be possible to verify, to a good accuracy, lepton universality for e-μ-τ.

## II. New physics predictions

At colliders, no significant deviations have been, so far, observed with respect to the SM. Only few **indications in B physics** appear in several channels, none of them having reached a 'discovery level' above 5 sd. The most intriguing feature of these deviations is that they seem to suggest violation of **lepton universality**. For instance the channels Kμμ, K*μμ, ϕμμ, D*τν behave differently than Kee,K*ee, ϕee, D*μν. This reminds us the long standing effect in g-2 with an anomaly only seen for the μ channel.

This apparent lack of lepton universality, if real, can be interpreted diversely and models based on leptoquarks or heavy Z' which couple non-universally have been suggested. These models seem to be bottom-up interpretations and one would rather favour a top-down approach which also addresses the main challenges of the SM, like the gauge hierarchy problem and the large mass hierarchy between fermions.

SUSY and the Randall Sundrum (RS) model address the **gauge hierarchy issue**. The latter [2] naturally predicts non-universal lepton couplings since it interprets the **hierarchy of fermion masses** as due to different locations of fermions in its extra dimension. In reference [3] one can find such an interpretation of the effects observed in B factories.

This state of affairs motivates examining lepton measurements at ILC250. While LEP1 has performed its mission by precisely measuring the lepton couplings to the Z boson, without observing any departure from universality, LEP2 has provided meagre accuracies in this domain [4]. To motivate Bhabha scattering measurements, I will examine the so-called **gauge-Higgs-universal** model (GHU), a version of RS which assumes that the Higgs boson is the fourth component of the vector mesons W± and Z, hence benefitting from the gauge symmetry protection to avoid the quadratic divergences affecting the Higgs mass in the SM, the so-called **naturality problem**. I will rely on the predictions given in [5].

In the GHU model, fermions of different chirality play different role and one expects that **right-handed fermions** are located near the Higgs brane and will therefore be coupled to the Kaluza Klein (KK) bosons of the RS model. This is in clear contrast to previous models where only heavy fermions were supposed to have such property. To test this model with a high sensitivity one should therefore use right-handed initial electrons, as provided by the ILC set up.

# III.  How to analyse the data?

## III.1 Inputs from LEP1

At the level of accuracy reachable by ILC250, one should worry about the experimental uncertainty on the Z couplings.

In the RS model, it is conceivable that these couplings could deviate from the SM due, e.g., to Z'-Z mixing. This idea has been used [6] to explain the origin of the deviation observed on the ZbRbR coupling at LEP1. No deviation is expected on electron couplings, within usual RS schemes where the electron is far from the Higgs brane while in the GHU model this would not be the case for the eR component which should be affected similarly to bR which is not compatible with LEP1 data on Zee. To avoid such contradiction in the GHU model, one needs to assume that there is no significant Z-Z' mixing and that the ZbRbR effect is either a statistical fluctuation or has another origin like a mixing effect between bR and a heavy quark b'R [7].

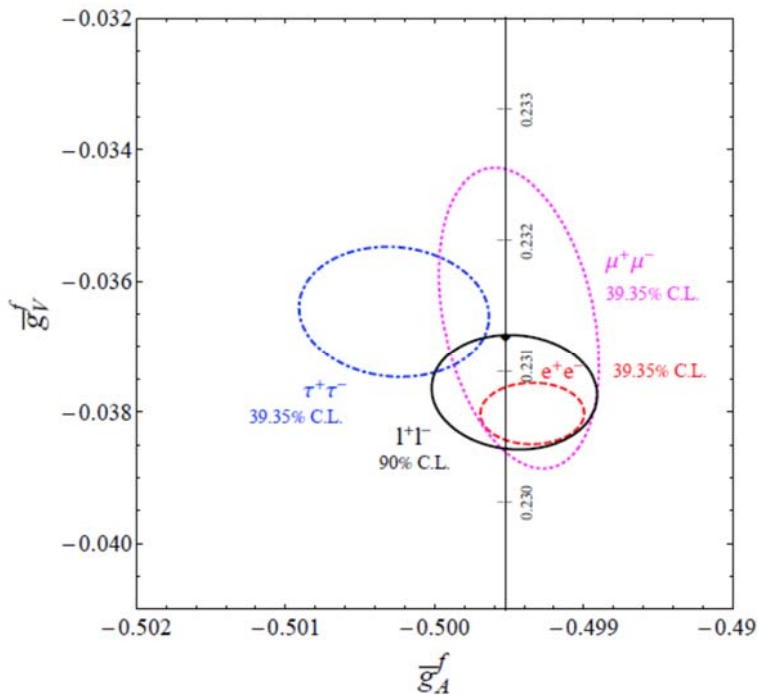

LEP1 provides a very precise measurement for lepton couplings, Zee being the most precise. There is no significant sign of departure from the SM as can be seen in figure 1 taken from the PDG [1]. Including the 3 lepton measurements, figure 1 gives **$s^2w=0.2308\pm0.0004$**. This value is compatible with the overall fit value which is more precise but could be differently affected by BSM physics. In the following I will stick to the **LEP1 determination** only.

Even with the large statistics accumulated at ILC250, Zee/µµ/ττ coupling measurements will not surpass LEP1. In other words, present errors on leptonic couplings given by LEP1 will provide a solid base line for the ILC measurements.

*Figure 1: LEP1 1 $\sigma$ ellipses for the measurement of Z couplings to ee, µµ and ττ. The dark contour corresponds to the 90% CL allowed region assuming lepton universality and diamond corresponds to SM best fit.*

We will see that they are so precise that they do not affect our conclusions. Note in passing that this is not the case for the channel ee->bb where, with beam polarisation and powerful b charge measurements, the ZbRbR measurement should surpass LEP1 [8].

---

[1] http://pdg.lbl.gov/2016/reviews/rpp2016-rev-standard-model.pdf

## III.2 The Bhabha channel

The Bhabha channel is particularly clean and efficiently reconstructed at ILC. This has been checked recently [9] where one has verified that the signal is essentially ~100% pure and can be detected with ~100% efficiency. It seems therefore straightforward to estimate the statistical accuracies that one can expect at ILC250 without using sophisticated simulation tools.

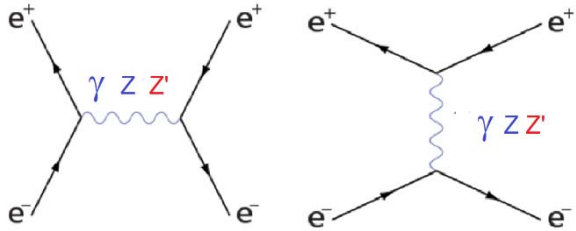

As illustrated by above diagrams, the complexity of this channel comes from the presence of a t-channel exchange due to the photon, the Z and, eventually, from new heavy bosons which are present in models beyond the SM.

For this analysis, I have used analytic formulae which can be found in [10]. They are at the Born level which is sufficient for present estimates. They include electron polarisation as for the SLC set up. I have verified that they agree with the formulae provided for LEP1 studies [11].

The cross section becomes very large in the forward hemisphere due to t-channel photon exchange and tends to dominate the cross section. In contrast to usual two fermion channels, one cannot trivially summarize the results of an analysis by just providing the total cross section and the forward-backward asymmetry AFBf which contain all the relevant information.

Z' exchange diagrams do not peak in the forward region and one could naïvely conclude that this channel has a poor sensitivity to new physics. This is however not the case given that the Z' exchange can interfere with the photon and the Z exchange. Moreover, the forward cross section is very large and therefore the statistical error is smaller than in the annihilation case.

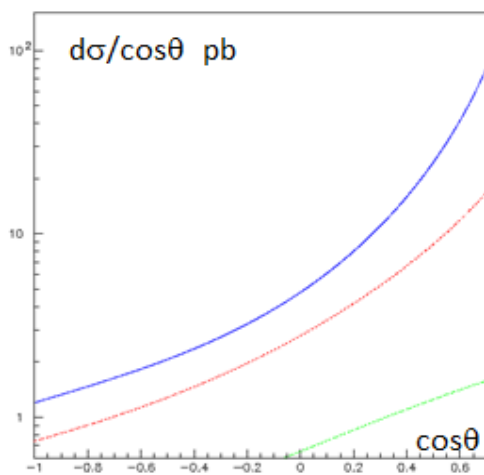

Figure 2: The 3 components of the Bhabha cross section at ILC250: photon t-channel exchange in blue, annihilation cross section in green, absolute value of interference Z-photon in red.

Figure 2 illustrates the composition of the SM components by displaying the angular distribution of the Bhabha scattering at ILC250. It shows that the resulting angular distribution is very much forward peaked due to the t-channel photon exchange (blue). It shows that the annihilation channel (green) is completely sub-dominant which seems to imply that any effect due to the exchange of a heavy boson will be lost in the background. This is not so, due to a strong interference between the photon exchange and the Z contribution as shown by the red curve. I will show in the GHU example that, due to this effect and due to the high rate provided by the photon exchange, far from being uninteresting the Bhabha process provides a sensitivity similar to μμ and ττ processes.

Generally speaking, the e+e- channel allows to unambiguously measure the vertex Z'ee which appears in all other annihilation channels, meaning that it is an essential measurement to fully disentangle BSM effects. The Zee vertex is very precisely known from LEP1 measurements as already mentioned. We therefore can safely extract the effect due to the Z' exchange. We will then be able to precisely verify lepton universality for the Z' by combining with the μμ and ττ channels which include the Zee vertex. Again, this type of test is by no means trivial since present B anomalies suggest that Z' could have a non-universal coupling to leptons, as naturally happens in RS models.

### III.3 LEP2 results

At LEP2, the data from the four collaborations were combined at various energies. The largest sample was collected at 189 GeV and figure 3 shows the result. The second plot shows how that this process is strongly peaked due to photon exchange. One is therefore unable to summarize the angular dependence through the usual indicator AFBf. Instead, one displays the ratio Data/SM as shown in the first plot. It reveals no anomaly [4].

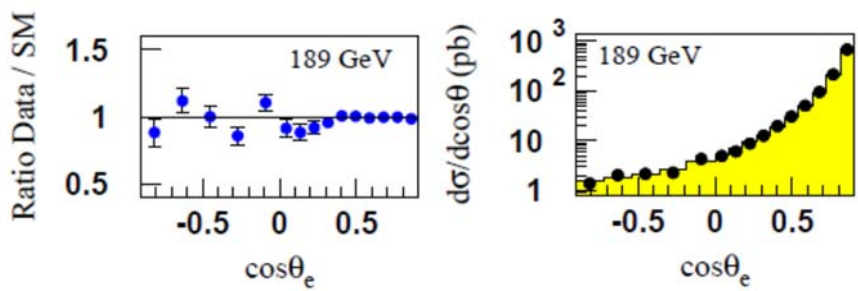

*Figure 3: Combined LEP2 data at 189 GeV. On the left is displayed the distribution of the ratio data/SM prediction vs cos$\theta$e, on the right the angular distribution of the cross section.*

Note in passing that, even for ordinary fermions, the AFBf vision is also too naïve since, for example, in the case of the top quark, electroweak corrections and non-resonant 6 fermion background tend to distort the naïve angular distribution at the Born level [12].

Here, instead, I will simply provide the predicted angular distribution, as given at Born level and show how it is affected by the presence of BSM physics by the defining the deviation to the SM. In the future, one can hope to have generators which contain all corrections relevant to the experimental accuracy.

### III.4 Luminosity and beam polarisation

The **high counting rate** allows to use the very forward region of our detector as a luminometer. In this region one expects that the photon exchange becomes completely dominant. We shall see in section IV how a **Z' exchange** can significantly, at the 0.1% level, influence the Bhabha cross section down to $\theta_e$~100 mrad. The luminometer [13] of ILD operates between 43 and 68 mrad, where the influence of this type of Z' is below 0.01%.

The uncertainty on the luminosity measurement comes mainly from two effects. First the geometry of the forward detector, which is not well enough determined. It comes from the transverse positioning of this calorimeter with respect to the beam which serves to define the scattering angle. Another source of uncertainty comes from an incomplete theoretical calculation of this process, insufficient to cope with the high statistical accuracy reachable at ILC250.

These aspects are relevant for the present analysis since the Z' effect remains significant in the forward region due to the very high statistical measurement accuracy.

In [13], a conservative estimate of the systematics is given, at the level of a few $10^{-3}$. If true, this would affect the present analysis which uses the forward region, forward down up to $\theta$ ~100 mrad or $\cos\theta=0.995$, where the statistical error is about $10^{-4}$ if one uses a $\cos\theta$ bin width of 0.1.

One can however object to this pessimistic conclusion.

First of all, the specific aspect of the Bhabha analysis is that it is embedded in the luminosity measurement, meaning that it is not used as an external information but is part of the measurement. It therefore does not suffer from the same systematic uncertainties. One should realize that BSM effects would change the **shape** of the angular distribution and cannot be absorbed just by changing the overall normalisation. Admittedly, it is preferable to keep such a correction at minimum to reach the best sensitivity in this measurement.

Second, as pointed out by the authors of [13], the figures quoted are conservative. For instance, concerning the small four fermion background, it seems plausible that it can be very precisely subtracted and corrected. Concerning self-focusing effect of the confronting beam, again it can be corrected, provided that one keeps tracks of the beam size variation with time.

The present analysis is, in part, intended to draw the attention on the importance of these aspects.

Not to be forgotten, is our need of an up to date generator, including corrections to allow reaching the $10^{-4}$ level of precision in the forward region, which requires a dedicated effort.

**Beam polarisation** is a very powerful tool to detect BSM physics. In the example given below, new physics mainly appears with eR polarisation. The uncertainty on this polarisation can be kept very small even if the positron beam is unpolarised, as could be the case at the start of the machine. One could use the process **ee->WW** as a powerful analyser to extract the beam polarisation since neutrino exchange only responds to eL.

At the moment [1], it is planned to favour eL in the sharing of luminosities given the slightly larger cross section for the ZH process, but this choice is certainly renegotiable and I will assume equal sharing, as for 500 GeV.

# IV. Expected effects for the GHU model

## IV.1 The GHU model

I use the GHU model to illustrate the power of this type of measurement. The reader is referred to reference [5] for the details. This reference provides all the couplings and masses of the heavy resonances needed to predict the deviations. The model depends on one parameter and 3 choices have been proposed. There are 3 massive resonances, two of them being simply γkk and Zkk, Kaluza Klein recurrences of the photon and Z boson, a third one being an extra resonances coming from an extended group (table 1). Table 2 shows that for leptons these heavy resonances are **primarily coupled to eR**, the right-handed electron, hence the power of polarized beams to increase by a factor ~2 the ratio signal/background. It also shows that there is a slight **departure from universality for the leptons**, challenging to measure.

If one decomposes the annihilation amplitude into the 3 independent chirality combinations LL, RR and RL (LR being identical in the case of ee), the dominant deviation corresponds to:

$$RR = Qe^2 + BW(-Qes^2w)^2/s^2wc^2w + \Sigma BW'k'(geR_{k'})^2/s^2w \text{ with } BW' = s/(s-m_{Z'}^2)$$

where $Qe=-1$ and where the coupling constants $geRk'$ are given in table 2. Below the resonance, the Z' part gives a negative contribution, BW'<0. LL and RL are almost SM like. Note that the right-handed coefficients in front of BW' are much larger than the SM Z couplings. For the 1$^{st}$ solution one has:

$$gRZ = -Qesw/cw = -0.55 \quad gR\gamma kk = -1.98/sw = -4.11 \quad gRZkk = 1.09/sw = 2.27 \quad gRZR = -1.50/sw = -3.12$$

On top of this, one has 3 resonances contributing, hence the expected high sensitivity.

| $\theta_H$ [rad.] | $zL/10^4$ | $m_{KK}$ [TeV] | $m_{Z(1)}$ [TeV] | $\Gamma_{Z(1)}$ [GeV] | $m_{\gamma(1)}$ [TeV] | $\Gamma_{\gamma(1)}$ [GeV] | $m_{Z_R}$ [TeV] | $\Gamma_{Z_R}$ [GeV] |
|---|---|---|---|---|---|---|---|---|
| 0.115 | 10 | 7.41 | 6.00 | 406 | 6.01 | 909 | 5.67 | 729 |
| 0.0917 | 3 | 8.81 | 7.19 | 467 | 7.20 | 992 | 6.74 | 853 |
| 0.0737 | 1 | 10.3 | 8.52 | 564 | 8.52 | 1068 | 7.92 | 1058 |

*Table 1: Masses and widths of Z' bosons*

3 set of parameters have been chosen, as seen from table 1. The couplings corresponding to the first set are shown in table 2. These couplings are almost the same for the 3 solutions.

| $f$ | $g^L_{Z^{(1)}f}$ | $g^R_{Z^{(1)}f}$ | $g^L_{Z_R^{(1)}f}$ | $g^R_{Z_R^{(1)}f}$ | $g^L_{\gamma^{(1)}f}$ | $g^R_{\gamma^{(1)}f}$ |
|---|---|---|---|---|---|---|
| $\nu_e$ | −0.1968 | 0 | 0 | 0 | 0 | 0 |
| $\nu_\mu$ | −0.1968 | 0 | 0 | 0 | 0 | 0 |
| $\nu_\tau$ | −0.1967 | 0 | 0 | 0 | 0 | 0 |
| $e$ | 0.1058 | 1.0924 | 0 | −1.501 | 0.1667 | −1.983 |
| $\mu$ | 0.1058 | 1.0261 | 0 | −1.420 | 0.1667 | −1.863 |
| $\tau$ | 0.1057 | 0.9732 | 0 | −1.354 | 0.1666 | −1.767 |
| $u$ | −0.1361 | −0.7152 | 0 | 0.9846 | −0.1111 | 1.2983 |
| $c$ | −0.1361 | −0.6631 | 0 | 0.9205 | −0.1111 | 1.2036 |
| $t$ | 0.5068 | −0.4764 | 1.0314 | 0.6899 | 0.4158 | 0.8666 |
| $d$ | 0.1665 | 0.3576 | 0 | −0.4923 | 0.05557 | −0.6491 |
| $s$ | 0.1664 | 0.3315 | 0 | −0.4602 | 0.05556 | −0.6018 |
| $b$ | −0.6303 | 0.2387 | 1.0292 | −0.3446 | −0.2082 | −0.4331 |

*Table 2: Couplings of neutral vector bosons (Z' bosons) to fermions in units of gw/sw for $\theta_H$=0.115.*

## IV.2 The GHU model at LHC

LHC searches for heavy bosons decaying into leptons provide constraints on this model [5], already excluding the first solution, based on ATLAS data collected in 2016. The absence of candidates with masses larger than 3 TeV is sufficient to do so. The reason for this precocious sensitivity to high Z' masses is not only due to large couplings but benefits from the very large width of the 3 resonances, which reflect the large couplings, as can be seen in table 1.

Recently [14] CMS has updated these searches with the 2016-17 data. In 2017, the search restricted to ee final states observes one candidate at ~3.5 TeV.

The progress for mass coverage with higher luminosity will be rather slow but, given the large width of this resonance and the absence of SM background, one can anticipate a coverage reaching ~8 TeV. In case one could double the energy with the HE-LHC option, this mass limit will also double.

## IV.3 The GHU model at ILC250

ILC250 will be able to confirm or rule out this model with a moderate fraction of the statistics as will be shown shortly. One could even wonder if LEP2 has already done so. This has been checked and the answer is negative. No wonder, since LEP2 has only collected 2.5 fb-1 without beam polarisation. The first figure below shows the angular dependence of the excess in % of the SM signal from its interference with the photon alone (full), and adding the Z-Z' interference (dotted). The second figure shows the angular dependence of the significance of the signal assuming purely statistical errors and a binning in $\cos\theta$ of 0.1. These results are shown for the 1$^{st}$ and 3$^{d}$ solution of [5].

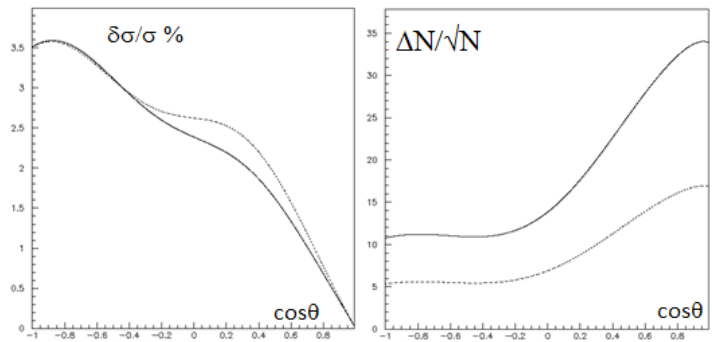

*Figure 4: On the left, the predicted excess with respect to the SM in % due to the Z' exchange. The full curve represents the term due to interference with the t-channel photon exchange; the dotted adds the interference Z-Z'. On the right, the expected statistical significance for a $\cos\theta$ bin width 0.1 of this excess at ILC250. The dotted curve corresponds to the third solution with mZ'=8.5 TeV.*

One sees that while the relative effect is the largest in the backward hemisphere, it significance is larger in the forward hemisphere. This illustrates the need to reduce systematical errors in the forward region $\cos\theta$~0.9, where the statistical error reaches 10$^{-4}$. For $\cos\theta$ ~0.9, one sees that the significance reaches its maximum, to more than 30 s.d. . This effect, with a signal divided by 15 remains significant and would correspond to a Z' mass of ~20 TeV.

One can also ask the following question: which luminosity is needed to observe a significant effect in the case of the 3d solution, the worse one? The answer is ~20 fb-1 with eR, that is ~2% of the total luminosity. This should happen in the early years of operation.

## IV.4 Annihilation channels

### IV.4.1 µµ and ττ

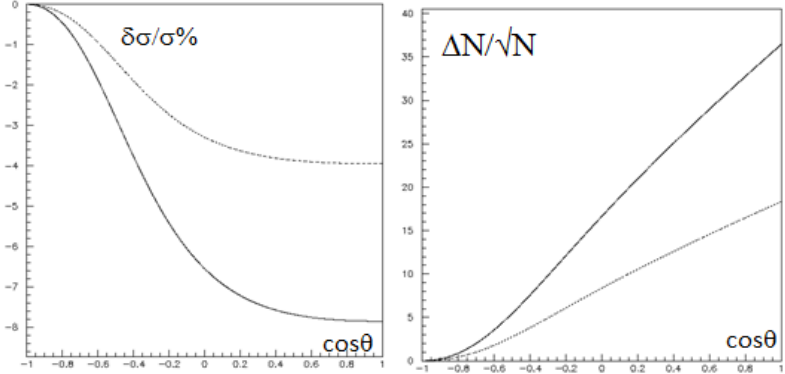

Figure 5: On the left, for mZ'=6 TeV, the predicted deficit in ee->µµ with respect to the SM in % due to the Z' exchange. The dotted curve corresponds to the solution with mZ'=8.5 TeV. On the right, the statistical significance of the effect assuming bins in $\cos\theta$ of 0.1. The dotted curves correspond to the solution with mZ'=8.5 TeV.

Above distributions can be simply interpreted. Recall that in the annihilation channel one has:

$$d\sigma R/d\cos\theta \sim (1+\cos^2\theta)(RR'^2+RL'^2)+2\cos\theta(RR'^2-RL'^2)$$

This distribution peaks at $\cos\theta=1$ where the cross section $\sim 4RR'^2$ is maximally affected by the Z' contribution, hence the maximum in significance observed in figure 5. Contrary to the ee->ee channel, the annihilation channels receive a negative contribution from the Z' exchange, since, as already mentioned, below the resonance one has BW'<0.

A similar effect is expected for the ττ channel.

### IV.4.2 bb and tt

For these modes, table 2 shows that, contrarily to the leptonic case where only RR' was affected, the RL' amplitude receives a large contribution.

In the b quark case, the selection efficiency for the cross section measurement is quite high, near 70%, leaving little background [15]. For the angular distribution measurement one needs further selections based on **b quark charge** measurement which leaves about 10% efficiency.

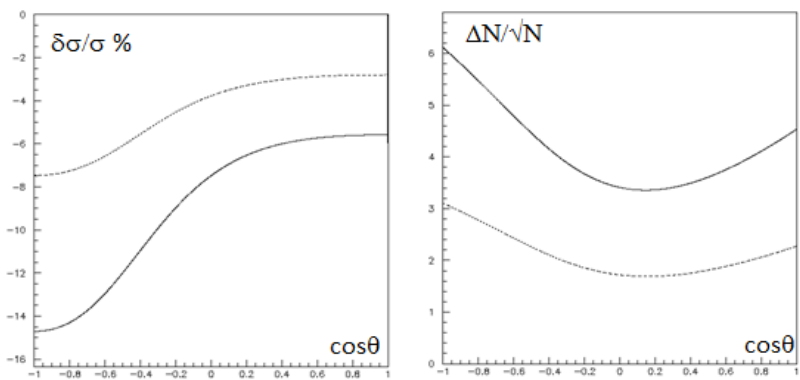

Figure 6: On the left, the predicted deficit in ee->bb with respect to the SM in % due to the Z' exchange, with mZ'=6 TeV. The dotted curve corresponds to the solution with mZ'=8.5 TeV. On the right, the statistical significance of the effect assuming $\cos\theta$ bins of 0.1. The dotted curves correspond to the solution with mZ'=8.5 TeV.

For the top quark, one needs to assume an operation at a centre of mass energy 500 GeV. It is expected that 2000 fb-1 will be collected with eR at this energy. The efficiency for the cross section measurement is also of order 70%. For the angular distribution measurement, one can use a combination of b quark charge measurements and t->b$\ell\nu$ decays, to reach an efficiency of ~30%.

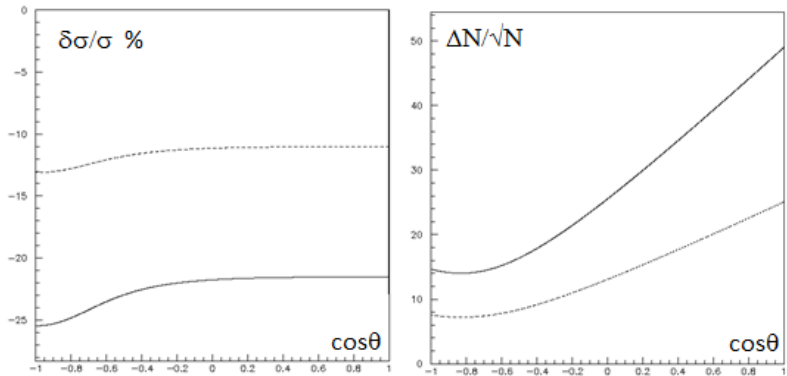

Figure 7: On the left, the predicted deficit in ee->tt with respect to the SM in % due to the Z' exchange for mZ'=6 TeV. The dotted curve corresponds to the solution with mZ'=8.5 TeV. On the right, the statistical significance of the effect assuming $\cos\theta$ bins of 0.1. The dotted curves correspond to the solution with mZ'=8.5 TeV.

Figure 7 displays an impressive sensitivity but one should not forget that benefits come from an increase the centre of mass of energy to 500 GeV, a doubling of the integrated luminosity, and an improved reconstruction efficiency with respect to the bb channel.

The following table summarizes the expected effects on the cross sections and the forward backward asymmetries for the channels under considerations. Recall that, for ee, a specific analysis is needed which uses the shape variation leaving free the overall normalisation.

From these quantities it is trivial to extract RR' and RL' and knowing the ee couplings, to derive separately the other couplings.

Table 3: Sensitivity of the various channels. The first error is statistical only. When a second error is given, it assumes an uncertainty on s²w from leptonic channel measurements at LEP1 of ±0.0004. The second efficiency line refers to the

| Channel c.o.m. energy | ee 250 GeV | μμ 250 GeV | bb 250 GeV | tt 500 GeV |
|---|---|---|---|---|
| Efficiency for σR % | 100 | 100 | 67 | 70 |
| δσR/σR %+δs²w | | -6.3±0.1±0.04 | -7.6±0.16±0.08 | -22.4±0.13±0.11 |
| MZ' limit TeV 90% CL | | >12 | >11.4 | >17.4 |
| Efficiency for cosθ % | | 100 | 10 | 20 |
| MZ' from cosθ TeV | >22 | >24 | >9 | >23 |

The cosθ **adjustment** gives the best result on Mz' limits, except for bb due to a reduced efficiency.

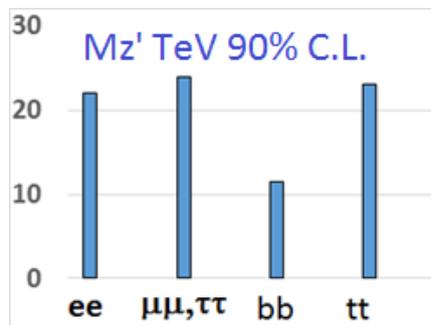

Table 3 shows that, for annihilation channels, the error due to the uncertainty on s²w is almost negligible except for what concerns the top quark measurement.

The tau result is not reported since it is almost identical to the μ channel. The deviation goes like BW' and is therefore reduced by 2 if one goes from 6 TeV to 8.5 TeV.

Figure 8: Mass coverage of ILC for the GHU model for various flavours. These limits correspond to ILC250 except the top case at 500 GeV.

Figure 8 shows that the GHU scenarios are covered beyond the requirements of [5].

Experimental errors do not take into account various systematical effects: from theory, experimental efficiency, background subtraction and luminosity. These will not significantly alter our conclusions

We show in section V how one should adapt such results for different models.

### IV.4.3 Comparisons

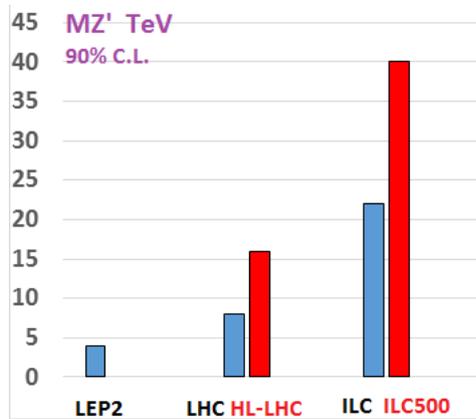

Figure 9: Comparison of the reach on GHU Z' mass for LEP2, LHC, HL-LHC and for ILC at 250 and 500 GeV.

One sees that, with beam polarisation, ILC will be able to unambiguously sort out the pattern predicted by this model and explore it up to masses of several tenths of TeV. Figure 9 summarizes the reach allowed for the GHU model. The Z' mass which is plotted corresponds to the average mass of the 3 heavy resonances of the model.

The GHU effect increases like s, meaning by 4 at 500 GeV. The statistical accuracy will be reduced by $\sqrt{2}$ since the cross section goes like s, an effect only partially compensated by an increase in luminosity of 2. All in all, the Z' mass limit reached for fermion channels will be increased by $2/2^{0.25}=1.68$. As an example the µµ channel will cover a mass range up to 40 TeV.

### IV.4.4 Z-Z' mixing?

One cannot a priori exclude that Zkk or ZR mix with Z and produce an observable effect at LEP1. From table 2, one sees that Zkk and ZR will couple preferentially to bL and therefore the resulting effect would be a larger deviation for ZbLbL, which goes against LEP1 observations. This means that the GHU model seems a priori excluded to explain the LEP1 effect on Zbb, if real. It needs an extension, e.g. considering a b' quark that mixes with b.

In [6], this effect is interpreted by Z-Z' mixing using a model where bR quarks are located near the Higgs brane, bL being away. For the electron part one predicts zero coupling to Z' and a minute, but non-zero coupling for Zkk and therefore almost no effect at LEP1. Above the resonance, γkk and Zkk will contribute negligibly to ee->ee but significantly for ee->bb and ee->tt.

In other words, the models [5] and [6] are well distinguished at ILC250.

## V. A model independent interpretation

Instead of a model dependent approach where the couplings are fixed, as for the GHU case, one can define a mass reach for ILC250 sensitivity with a model independent approach. Here I will simply give the recipe to deduce a generic Z' mass limit from the GHU limits given in table 3.

In GHU, where the 3 resonances are almost degenerate in mass, one can define an equivalent effect as a single Z' with a coupling gGHU such that:

$\Sigma BW_{k'}(geR_{k'})^2/s^2w=(gGHU)^2 BWZ'$   which gives $(gGHU)^2=(-4.11)^2+(2.27)^2+(-3.12)^2=(5.6)^2$

From which one deduces that:

$$M_{Z'} = M_{GHU}(g_{eRZ'}/g_{GHU})$$

The gZ' coupling can vary over a large range. gZ' can be below 1, if Z' would be coupled to μμ as the SM Z boson, gμRZ'~0.55. Then the 24 TeV limit for μμ would melt into 2.4 TeV which is already excluded by LHC. gZ' can be as high as $4\pi$, in which case ILC250 would cover masses up to 50 TeV.

These examples provides a clear illustration of the merits of ILC250 as compared to LHC. For large couplings, as in the GHU case, ILC250 covers masses way beyond the LHC domain, while for small couplings, LHC has a superior reach.

## VI. Test of lepton universality

To test μ/e universality, one can measure the ratio Rμ=ee->Z'->μμ/ee->Z'->ee. The same is true for the ττ channel. This quantity measures the ratio of the right-handed couplings to the square. Any departure from 1 of this ratio would sign an effect of non-universal coupling to Z'. The error on this ratio depends on the intensity of the observed deviation, hence on the Z' mass. If one assumes Mz'=6 TeV, one finds an error of ~3% for Rμ, from which universality at the level of couplings can be tested with an accuracy of ~2%. Table 2 tells us that e and μ couplings differ by about 6%. For the τ lepton, the τ/e couplings differ by 12% with the same accuracy measurement of 2%.

In most other RS theories, geRZkk is expected to be small (although not negligible), meaning that any deviation observed in μμ or ττ and not seen in ee would prove non-universality. In all cases, one expects that the predominant deviation should occur in ττ.

## SUMMARY and CONCLUSIONS

The **sensitivity** of the e+e->e+e- reaction to BSM physics mainly comes from the interference of the Z' exchange with the t-channel photon exchange. This sensitivity mainly originates from the forward region due to high rate. This feature puts some pressure on the detector since **instrumental errors** need to be kept at a minute level. One would certainly benefit from an **absolute luminosity** measurement at the $10^{-4}$ level, which is considered quite challenging, but this accuracy is not mandatory since the effect of a Z' exchange is **angular dependent** and cannot be absorbed in a simple normalisation factor.

Requirements on HO **corrections for the generator,** are required at the same level, in particular for effects which can affect the shape of the **angular distribution**.

The couplings of a Z' to e+e- could be measured at ILC250 with high accuracy, up to Z' masses well **beyond LHC** in the case of the **GHU model**. In models where the Z' couplings are helicity dependent, as is the case for the GHU model, **beam longitudinal polarisation** increases the ratio s/b.

In any case, measuring the e+e- coupling is of fundamental importance to allow disentangling any deviation observed in the fermionic sector. In particular, such a measurement allows **testing GHU universality of the lepton couplings** at the % level.

For **heavy quarks**, large deviations are also expected, in particular for the top channel, assuming a 500 GeV centre of mass energy.

GHU predicts a very specific pattern of deviations for the fermionic couplings, unambiguously identified at ILC.

These Z' mass limits can be translated in any BSM model given the coupling constants.

The present note is meant to stimulate ongoing efforts for the preparation of the ILC experiments for what concerns the **Bhabha channel**.

**Acknowledgements.** The present work has benefitted from useful discussions with my colleagues Roman Poeschl and Marcel Vos. Marcel has attracted my attention on the need to constrain the e+e- couplings for EFT analyses involving the top quark measurements. For the GHU model, enlightening discussions with Yutaka Hosotani are gratefully acknowledged.

**References**

[1] Physics Case for the 250 GeV Stage of the International Linear Collider
Keisuke Fujii et al. . Oct 20, 2017.
e-Print: arXiv:1710.07621
The International Linear Collider Technical Design Report -
Volume 3.II: Accelerator Baseline Design Chris Adolphsen (ed.) et al.. Jun 26, 2013.
arXiv:1306.6328
The International Linear Collider Machine Staging Report 2017
Lyn Evans, Shinichiro Michizono. Nov 1, 2017.
e-Print: arXiv:1711.00568
[2] A Large mass hierarchy from a small extra dimension
Lisa Randall, Raman Sundrum. May 1999.
Published in Phys.Rev.Lett. 83 (1999) 3370-3373
e-Print: hep-ph/9905221
[3]Warped Electroweak Breaking Without Custodial Symmetry
Joan A. Cabrer, Gero von Gersdorff, Mariano Quiros. Nov 2010.
Published in Phys.Lett. B697 (2011) 208-214
e-Print: arXiv:1011.2205
[4]Electroweak Measurements in Electron-Positron Collisions at W-Boson-Pair Energies at LEP
ALEPH and DELPHI and L3 and OPAL and LEP Electroweak Collaborations (S. Schael et al.). Feb 14, 2013.
Published in Phys.Rept. 532 (2013) 119-244
e-Print: arXiv:1302.3415
[5] Distinct signals of the gauge-Higgs unification in e+e− collider experiments
Shuichiro Funatsu, Hisaki Hatanaka, Yutaka Hosotani, Yuta Orikasa. May 15, 2017.
Published in Phys.Lett. B775 (2017) 297-302
e-Print: arXiv:1705.05282
[6]Resolving the A(FB)**b puzzle in an extra dimensional model with an extended gauge structure
Abdelhak Djouadi, Gregory Moreau, Francois Richard. Oct 2006. 23 pp.
Published in Nucl.Phys. B773 (2007) 43-64
e-Print: hep-ph/0610173

[7] Vector-like top/bottom quark partners and Higgs physics at the LHC
Andrei Angelescu, Abdelhak Djouadi, Grégory Moreau. Oct 26, 2015. 28 pp.
Published in Eur.Phys.J. C76 (2016) no.2, 99
LPT-ORSAY-15-74
e-Print: arXiv:1510.07527

[8] Measurement of b quark EW couplings at ILC
S. Bilokin, R. Pöschl, F. Richard. Sep 13, 2017.
e-Print: arXiv:1709.04289

[9] Study of fermion pair productions at the ILC with center-of-mass energy of 250 GeV
Hiroaki Yamashiro, Kiyotomo Kawagoe, Taikan Suehara, Tamaki Yoshioka, Keisuke Fujii, Akiya Miyamoto. Jan 15, 2018.
e-Print: arXiv:1801.04671

[10] A Measurement of the effective electron neutral current coupling parameters from polarized Bhabha scattering at the Z0 resonance
Matthew D. Langston (Oregon U. & SLAC). Jun 2003.
UMI-30-95258, SLAC-R-629, SLAC-R-0629, SLAC-629, SLAC-0629, UMI-30-95258-MC
http://www.slac.stanford.edu/pubs/slacreports/reports03/slac-r-629.pdf

[11] LEP studies CERN 89-08
Bhabha scattering (p. 171)
http://cds.cern.ch/record/116932/files/CERN-89-08-V-1.pdf

[12] Probing New Physics using top quark polarization in the e+e- -> t \bar{t} process at future Linear Colliders
P.H. Khiem, E. Kou, Y. Kurihara, F. Le Diberder. Mar 13, 2015. 14 pp.
e-Print: arXiv:1503.04247

[13] Luminosity Measurement at ILC
I. Bozovic-Jelisavcic, H. Abramowicz, P. Bambade, T. Jovin, M. Pandurovic, B. Pawlik, C. Rimbault, I. Sadeh, Ivan Smiljanic. Jun 2010.
e-Print: arXiv:1006.2539

[14] Search for high-mass resonances in dilepton final states in proton-proton collisions at s√= 13 TeV
CMS Collaboration (Albert M Sirunyan et al.). Mar 16, 2018.
e-Print: arXiv:1803.06292
Search for high mass resonances in dielectron final state
CMS Collaboration. Mar 12, 2018.
CMS-PAS-EXO-18-006

[15] Sviatoslav Bilokin, PhD Thesis. July 18 2017.
"Hadronic showers in a highly granular silicon-tungsten calorimeter and production of bottom and top quark at the ILC"
LAL-17-050.

[16] A precise characterisation of the top quark electro-weak vertices at the ILC
M.S. Amjad et al.. May 22, 2015.
Published in Eur.Phys.J. C75 (2015) no.10, 512
e-Print: arXiv:1505.06020